# Universal Synchronous Spin Rotators for Electron-Ion Colliders

P. Chevtsov
*PSI, 5232 Villigen, Switzerland*
Y. Derbenev, G. Krafft, Y. Zhang
*Jefferson Lab, Newport News, VA 23606, USA*

The paper provides mathematics and physics considerations concerning a special class of electron spin manipulating structures for future Electron-Ion Collider (EIC) projects. These structures, which we call Universal Synchronous Spin Rotators (USSR), consist of a sequence of standard basic spin manipulating elements or cells built with two solenoids and one bending magnet between them. When integrated into the ring arcs, USSR structures do not affect the central particle orbit, and their spin transformation functions can be described by a linear mathematical model. In spite of being relatively simple, the model allows one to design spin rotators, which are able to perform spin direction changes from vertical to longitudinal and vice versa in significant continuous intervals of the electron energy. This makes USSR especially valuable tools for EIC nuclear physics experiments.

## 1. INTRODUCTION

The electron ring of JLEIC (a former MEIC), the Jefferson Lab's design of the Electron Ion Collider [1], was developed to most efficiently preserve and manipulate highly polarized electron beams for nuclear physics research. The electron polarization is vertical in the arcs to minimize (or maximize, if needed) depolarization effects, and longitudinal at the collision points as required by experiments (see Figure 1.1. below).

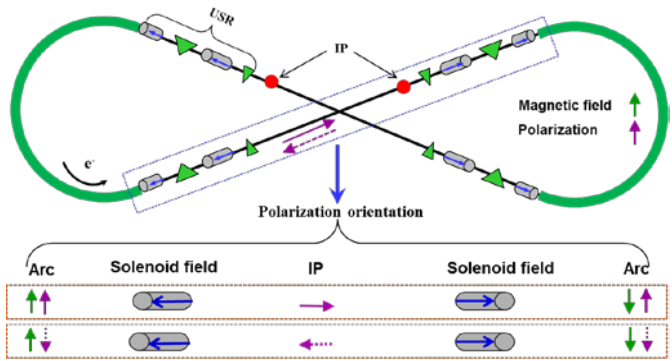

Figure 1.1: JLEIC electron polarization ($\vec{S}$, a purple arrow) directions remain the same in two arcs by having opposite longitudinal solenoid field directions in the same long straight. The blue arrow in a solenoid represents the field direction. The polarization orientation in one of the two long straights (shown in the blue dash-line box) is exhibited under the half brace with two different polarization states at the IP.

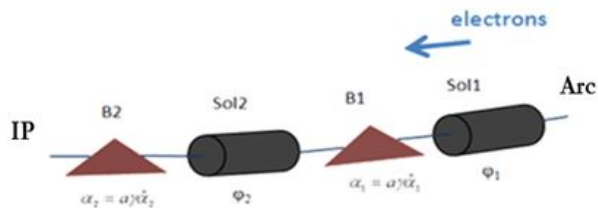

Figure 1.2: Universal Spin Rotator (USR) elements.

The proper spin orientation is accomplished using four Universal Spin Rotators (USR) located at each end of two arcs. These rotators, composed of interleaved solenoid and dipole fields as shown schematically in Figure 1.2 above, are designed to manipulte the electron polarization in the entire projected energy range from 3 to 12 GeV. The transverse orbital coupling induced by the longitudinal fields in solenoids is neutralized by placing quadrupoles between half solenoids [1].

We note that some kinds of spin rotators, which are also based on two bends and two solenoids but designed for beam extraction lines, were proposed in several publications including, for example, [3].

The electron polarization configuration, combined with a figure-8 geometry of the collider ring, produces a net zero spin precession. Hence the spin tune on the design orbit is zero and independent of the beam energy. This significantly reduces the synchrotron sideband resonances. In addition, since there is no preferred electron spin direction, the beam polarization can be easily controlled and stabilized by using relatively small magnetic fields such as, for example, spin tuning solenoids in the straights where the polarization is longitudinal.

We note that the USR is a simplest representative of the whole class of universal electron spin manipulating structures suitable for the JLEIC project. In this paper, we introduce this class and demonstrate its performance features. The most interesting fact is that from the mathematical point of view such universal spin rotating structures can fully be described by the elementary ("high school") trigonometry. Of course, it goes without saying that the physics of electron spin dynamics in accelerators must always be kept in mind.

All our considerations will be done in a dedicated Cartesian coordinate system (xyz), which we call a spin rotator definition coordinate system (see Fig. 1.3 below).

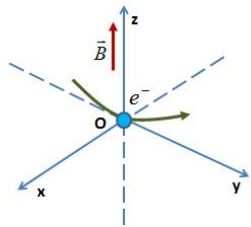

Figure 1.3: Spin rotator definition coordinate system.

The system is moving along the JLEIC ring together with the electron beam and is characterized by the next properties.

1. It is defined in the ring arc where the beam guiding magnetic field is directed upwards with respect to the surface of the Earth. We call this arc the principal spin rotator definition arc. At that, the other ring arc acts as a complementary one.
2. The axis Oz points in the direction of the guiding magnetic field.
3. The axis Oy follows the direction of the particle motion.
4. Finally, the axis Ox is oriented to form a right hand system (xyz).

Concentrate now on the principal spin rotator definition arc and introduce the next three matrices, which describe the spin vector rotations by the angles:

- $\varphi$ around the axis Oy

$$ROT(Oy,\varphi) = \begin{pmatrix} \cos\varphi & 0 & \sin\varphi \\ 0 & 1 & 0 \\ -\sin\varphi & 0 & \cos\varphi \end{pmatrix} \quad (1.1)$$

- $\alpha$ around the axis Oz

$$ROT(Oz,\alpha) = \begin{pmatrix} \cos\alpha & -\sin\alpha & 0 \\ \sin\alpha & \cos\alpha & 0 \\ 0 & 0 & 1 \end{pmatrix} \quad (1.2)$$

and

- $\theta$ around the axis Ox

$$ROT(Ox,\theta) = \begin{pmatrix} 1 & 0 & 0 \\ 0 & \cos\theta & -\sin\theta \\ 0 & \sin\theta & \cos\theta \end{pmatrix} \quad (1.3)$$

respectively.

By convention, positive rotations go counter-clockwise and negative rotations go clockwise. We note that in this paper we consider only electrons moving along the central beam trajectory. All typical beam optics and spin dynamics problems associated with various deviations of particles from this trajectory are planned to be analyzed in separate publications.

## 2. UNIVERSAL SYNCHRONOUS SPIN ROTATORS

In this chapter, we introduce a class of electron spin manipulating structures, which we call Universal Synchronous Spin Rotators (USSR).

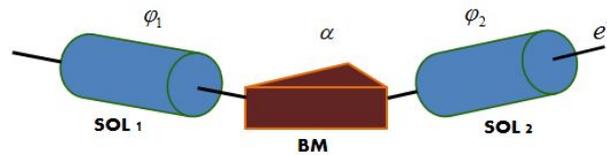

Figure 2.1: Elements of a Universal Synchronous Spin Rotator cell.

The USSR consists of several (two or more) standard cells. Each USSR cell (see Fig. 2.1 above) is built with two solenoids (SOL 1 and SOL 2) and one bending magnet (BM) between them. The BM is assumed to be a part of the machine ring, which makes such cells not affecting the central electron beam trajectory.

We require that each USSR cell operates exclusively in the vertical plane (yz), which means that

- the electron spin vector is initially located in (yz);

- the main function of the USSR cell is to rotate this vector through the angle $\varphi_1$ around the axis Oy before the BM and then to always bring this vector back into (yz) after the magnet by rotating it through the angle $\varphi_2$ around the same axis Oy.

We also require that the absolute value of the electron spin rotation angle $\alpha$ in the BM of any USSR cell does not exceed $2\pi$:

$$0 < |\alpha| \leq 2\pi \quad (2.1)$$

which reflects a practical range of electron spin rotation angles that can be achieved in JLEIC conditions with the use of such spin manipulation structures.

Recall that

$$\alpha = a \cdot \gamma \cdot \hat{\alpha} \quad (2.1.1)$$

where $a$ is the anomalous magnetic moment ($a \approx 0.00115965$), $\gamma$ is the relativistic factor, and $\hat{\alpha}$ is the bending angle of the electron moving in the magnetic field of the BM.

In addition, it is natural to limit spin rotations in USSR solenoids by the angles

$$-\pi \leq \varphi_1 \leq \pi \ , \quad -\pi \leq \varphi_2 \leq \pi \quad (2.2)$$

and to distinguish between the lower ($-\pi \leq \theta \leq 0$) and the upper ($0 \leq \theta \leq \pi$) half of the vertical plane (yz), which we denote as [yz, D] and [yz, U] respectively.

Write now the cumulative spin rotation matrix $M_c$ of the USSR cell:

$$M_c = ROT(Oy, \varphi_2) \cdot ROT(Oz, \alpha) \cdot ROT(Oy, \varphi_1) \quad (2.3)$$

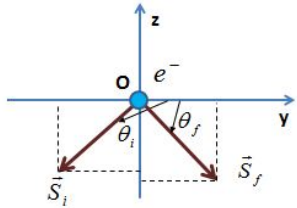

Figure 2.2: Spin vector transformation performed by a USSR cell.

So, if the electron spin vector $\vec{S}$ has the angle $\theta_i$ ("initial") with the axis Oy at the entrance into the USSR cell and the angle $\theta_f$ ("final") with the same axis at the exit from this cell (see Fig. 2.2 above), then the resultant spin vector transformation from $\vec{S}_i = (0, \cos\theta_i, \sin\theta_i)$ to $\vec{S}_f = (0, \cos\theta_f, \sin\theta_f)$ can be expressed in the next matrix form:

$$\vec{S}_f = M_c \cdot \vec{S}_i$$

or, more explicitly, as

$$\begin{pmatrix} 0 \\ \cos\theta_f \\ \sin\theta_f \end{pmatrix} = \begin{pmatrix} \cos\alpha\cos\varphi_2\cos\varphi_1 - \sin\varphi_2\sin\varphi_1 & -\sin\alpha\cos\varphi_2 & \cos\alpha\cos\varphi_2\sin\varphi_1 + \sin\varphi_2\cos\varphi_1 \\ \sin\alpha\cos\varphi_1 & \cos\alpha & \sin\alpha\sin\varphi_1 \\ -\cos\alpha\sin\varphi_2\cos\varphi_1 - \cos\varphi_2\sin\varphi_1 & \sin\alpha\sin\varphi_2 & -\cos\alpha\sin\varphi_2\sin\varphi_1 + \cos\varphi_2\cos\varphi_1 \end{pmatrix} \cdot \begin{pmatrix} 0 \\ \cos\theta_i \\ \sin\theta_i \end{pmatrix} \quad (2.4)$$

The last equation results in the following relations to be fulfilled:

$$\sin\varphi_1 = \frac{\cos\theta_f - \cos\alpha\cos\theta_i}{\sin\alpha\sin\theta_i}, \quad \sin\alpha\sin\theta_i \neq 0 \quad (2.5)$$

$$\sin(\varphi_2 + \chi) = 0, \quad \tan\chi = \frac{-\sin\alpha\cos\theta_i + \cos\alpha\sin\varphi_1\sin\theta_i}{\cos\varphi_1\sin\theta_i}$$

It is easy to see that the most important special cases in the above relations are associated with the equality

$$\sin\alpha\sin\theta_i = 0 \quad (2.5.1)$$

and correspond to the next three obvious situations.

S1. The initial spin vector direction is parallel $(\theta_i = 0)$ or antiparallel $(\theta_i = \pm\pi)$ to the direction of the particle motion (axis Oy).

S2. The spin vector makes in the BM a complete turn ($\alpha = \pm 2\pi$) or half a turn $(\alpha = \pm\pi)$ around the vertical axis Oz.

S3. The combination of S1 and S2.

In all these situations, one can obtain the desired parameters of the USSR cell directly from its original model (2.4). Some results are shown here in Fig. 2.3-2.4 below.

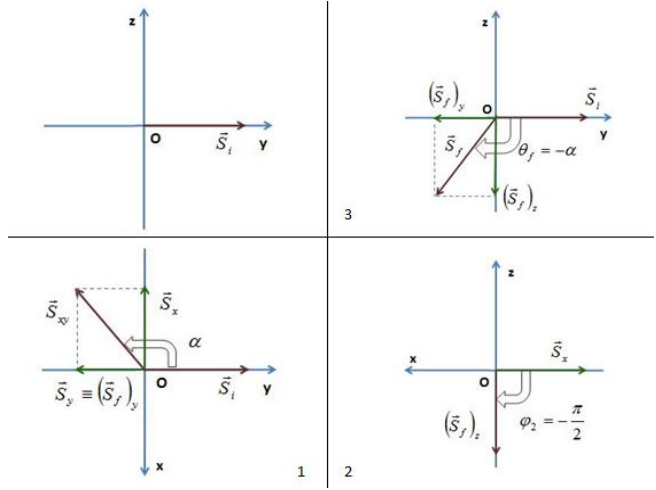

Figure 2.3: Spin vector transformation steps (1-3) in a USSR cell when the initial spin vector is parallel to the axis Oy. The spin rotation angle in the first solenoid $\varphi_1$ can be any value.

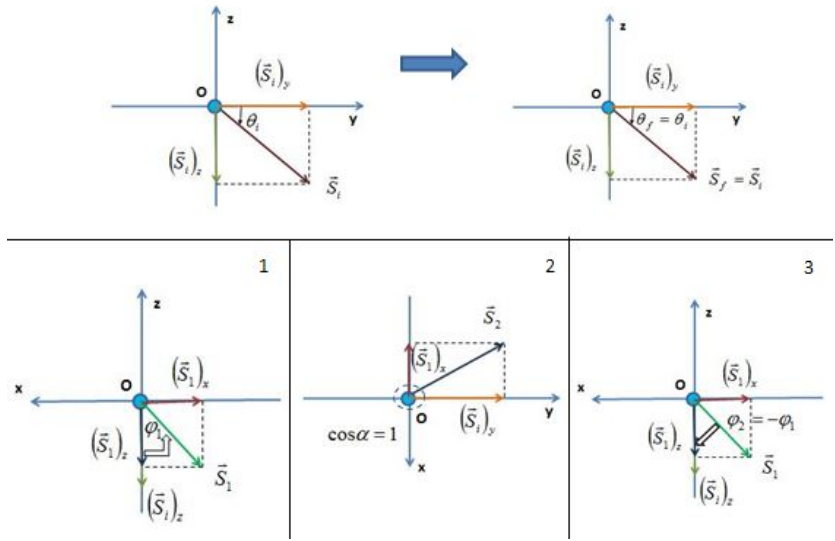

Figure 2.4: Spin vector transformation steps (1-3) in a USSR cell when this vector makes a complete turn (step 2) in the BM. Again, the spin rotation angle in the first solenoid $\varphi_1$ can be any value.

Return now to the main equation in relations (2.5)

$$\sin\varphi_1 = \frac{\cos\theta_f - \cos\alpha\cos\theta_i}{\sin\alpha\sin\theta_i}, \quad \sin\alpha\sin\theta_i \neq 0 \tag{2.6}$$

We see that spin rotation angles in the USSR cell cannot be arbitrary. In order to serve properly, the electron spin rotation angle in the DM together with the initial and final spin vector orientation angles must satisfy the next fundamental USSR condition:

$$-1 \leq \left( \frac{\cos\theta_f - \cos\alpha \cos\theta_i}{\sin\alpha \sin\theta_i} \right) \leq 1 , \quad \sin\alpha \sin\theta_i \neq 0 \tag{2.7}$$

This condition, combined with special cases mentioned above, is equivalent to the next two equation systems:

$$\cos(\alpha - \theta_i) \leq \cos\theta_f \leq \cos(\alpha + \theta_i) \tag{2.8}$$

$$\sin\alpha \sin\theta_i \leq 0 \tag{2.8.1}$$

and

$$\cos(\alpha + \theta_i) \leq \cos\theta_f \leq \cos(\alpha - \theta_i) \tag{2.9}$$

$$\sin\alpha \sin\theta_i \geq 0 \tag{2.9.1}$$

Now, if we have a set of $N$ standard USSR cells enumerated by the index $k$ $(k = 1,...N)$, then the total USSR is defined by the following equation systems:

$$\begin{array}{ll} \cos(\alpha_k - \theta_{i,k}) \leq \cos\theta_{f,k} \leq \cos(\alpha_k + \theta_{i,k}) & \cos(\alpha_k + \theta_{i,k}) \leq \cos\theta_{f,k} \leq \cos(\alpha_k - \theta_{i,k}) \\ \sin\alpha_k \cdot \sin\theta_{i,k} \leq 0 & \sin\alpha_k \cdot \sin\theta_{i,k} \geq 0 \end{array} \tag{2.10}$$

Here $\theta_{i,1}$ is a constant angle describing the electron spin direction at the entrance into the rotator and $\theta_{i,k+1} \equiv \theta_{f,k}$ $(k = 1,...N-1)$.

From the above statements immediately follows that the cells of the operational USSR must work synchronously, hence the name.

In case of a JLEIC flat electron ring, the most important electron spin original $\vec{S}_0$ (at the entrance into the USSR) and final $\vec{S}_F$ (at the exit from the USSR) directions are mutually exclusive vertical (along the axis Oz) and longitudinal (along the axis Oy). So, if we denote vertical spin directions as "d" (downwards) and "u" (upwards), as well as longitudinal spin directions as "p" (parallel to the particle motion direction) and "a" (anti-parallel to the particle motion direction) then we get two mutually exclusive basic spin direction pairs: ("d", "u") and ("p", "a"), which we call complimentary to each other.

At this point, it should be noted that in accordance with relations (2.5), one can always move the electron spin vector in any particular USSR cell from [yz, D] to [yz, U] and vice versa by taking

$$\varphi_2 = \pi - \chi \quad \text{if} \quad \chi > 0$$

and $\hspace{8cm} (2.11)$

$$\varphi_2 = -\pi - \chi \quad \text{if} \quad \chi < 0$$

instead of its basic value $\varphi_2 = -\chi$. At that, the value of $\varphi_1$ doesn't depend on whether the final spin vector in that cell belongs to [yz, D] or [yz, U]. This property defines a specific USSR (xy)-mirror symmetry.

Introduce the next USSR types, which we call USSR(I1, I2, I3) and which are characterized by the next meanings of indices I1, I2, and I3:

- I1 describing the electron spin direction at the entrance into the first USSR cell is any value from the basic spin directions ("d", "u") and ("p", "a");
- I2 is either "+" if electrons rotate counter-clockwise in USSR BM or "–" if they do it clockwise;
- I3 describing the electron spin direction at the exit from the last USSR cell is any value from the basic spin direction pair, which is complementary to the pair consisting of the value I1.

We also require that a USSR, which has the index "d" in its name consists of cells functioning exclusively in the [yz, D] half-plane. Similarly, a USSR having the index "u" into its name serves exclusively in the [yz, U] half-plane.

At this point, we note that the introduced USSR(I1, I2, I3) types are complementary in a sense that can easily be converted into each other by the proper elementary transformations of their basic working parameters. More details will be provided later in this paper.

Based on this definition, everywhere further in this paper, if it is not otherwise specified, we will assume that the USSR type is USSR(d, +, p).

So, each USSR cell is characterized by the next conditions to be fulfilled:

$$\cos(\alpha - \theta_i) \leq \cos\theta_f \leq \cos(\alpha + \theta_i)$$
$$0 \leq \alpha \leq \pi \qquad (2.12)$$
$$-\pi \leq \theta_i \leq 0$$

and

$$\cos(\alpha + \theta_i) \leq \cos\theta_f \leq \cos(\alpha - \theta_i)$$
$$\pi \leq \alpha \leq 2\pi \qquad (2.13)$$
$$-\pi \leq \theta_i \leq 0$$

As a result, the total USSR consisting of $N$ standard USSR cells is defined by the next $N$ equation systems:

$$\cos(\alpha_k - \theta_{i,k}) \leq \cos\theta_{f,k} \leq \cos(\alpha_k + \theta_{i,k}) \qquad \cos(\alpha_k + \theta_{i,k}) \leq \cos\theta_{f,k} \leq \cos(\alpha_k - \theta_{i,k})$$
$$0 \leq \alpha_k \leq \pi \qquad\qquad\qquad\qquad \pi \leq \alpha_k \leq 2\pi \qquad (2.14)$$
$$-\pi \leq \theta_{i,k} \leq 0 \qquad\qquad\qquad\qquad -\pi \leq \theta_{i,k} \leq 0$$

in which $\theta_{i,1} = -\dfrac{\pi}{2}$, $\theta_{f,N} = 0$, $\theta_{i,k+1} = \theta_{f,k}$ $(k = 1,...N-1)$.

Recall that from the USSR definition follows that electron spin rotation angles in its dipole magnets (as parts of the ring) are always proportional to each other. At the same time, the equations (2.14) determine the dependence of $\theta_{f,k}$ on $\theta_{i,k}$ and $\alpha_k$.

In this paper, we restrict our considerations to those USSR, which are characterized by linear relations between spin rotation angles in standard cells:

$$\theta_{f,k} = \hat{m}_k \cdot \alpha_k + \hat{b}_k \qquad (2.16)$$

where $\hat{m}_k$ and $\hat{b}_k$ are constant parameters.

We call the value of $\left|\hat{b}_k\right|$ the correcting phase of the cell number $k$.

### 3. SEQUENTIALLY DECREASING CORRECTING PHASE USSR

Introduce now a group of linear USSR, which are called Sequentially Decreasing Correcting Phase USSR or, in short, SDCP USSR.

The group is characterized by the next properties.

P1. All USSR cells, except the last one, are equipped with equal BM:

$$\alpha_1 = \alpha_2 = ... = \alpha_{N-1} \equiv \alpha \qquad (3.1.1)$$

P2. The BM of the last cell is two times weaker:

$$\alpha_N = \frac{1}{2}\alpha \qquad (3.1.2)$$

P3. The spin rotation angles in USSR cells are connected by relations:

$$\theta_{f,k} = -m_k \cdot \alpha_k - b_k \qquad (3.2)$$

in which

$$k = 1,...N-1$$

$$b_k = \frac{\pi}{2} \cdot \left(1 - \frac{k}{N-1}\right) \qquad (3.3.1)$$

$$m_k = \frac{1}{2} - \frac{b_k}{\pi} = \frac{k}{2 \cdot (N-1)} \qquad (3.3.2)$$

It is worth mentioning that the properties P1-P3

- automatically provide one of the most important requirements to the operational USSR, which is matching of $\theta_{f,N-1}$ to the spin rotation angle in the last USSR cell

and

- with the definition

$$b_0 \equiv -\theta_{i,1} \tag{3.4}$$

naturally include the initial condition $\theta_{i,1} = -\dfrac{\pi}{2}$.

It appears that SDCP USSR structures are extremely important for JLEIC electron spin manipulation tasks. In particular, they allow one to transform the electron spin vector direction from original $\vec{S}_0 = \vec{S}_{i,1}$ to final $\vec{S}_F = \vec{S}_{f,N}$ for a substantially large ratio of electron energies

$$(\Delta E_{USSR,N})_r = \frac{E_{USSR,N,\max}}{E_{USSR,N,\min}} \tag{3.5}$$

where $E_{USSR,N,\min}$ and $E_{USSR,N,\max}$ define a continuous energy interval

$$(\Delta E_{USSR,N}) = [E_{USSR,N,\min}, E_{USSR,N,\max}] \tag{3.6}$$

covered by the structures.

We note that in the geometrical sense each individual USSR cell number $k$ naturally limits such a ratio

$$(\Delta E_k)_r = \frac{E_{k,\max}}{E_{k,\min}} = \frac{\alpha_{k,\max}}{\alpha_{k,\min}} \tag{3.7}$$

by the angles $\alpha_k \in [\alpha_{k,\min}, \alpha_{k,\max}]$, for which the corresponding curve $\cos\theta_{f,k}$ doesn't go outside the area between $\cos(\alpha_k - \theta_{i,k})$ and $\cos(\alpha_k + \theta_{i,k})$, in accordance with equations (2.1.1) and (2.14).

One of the most remarkable features of SDCP USSR with $N$ cells is that for all $k = 1,...N-1$

$$\alpha_{k,\min} = \frac{\pi}{(2N-1)} \tag{3.8}$$

and

$$\alpha_{k,\max} = \frac{\pi}{(2N-1)} \cdot (4N-3) = (4N-3) \cdot \alpha_{k,\min} \tag{3.9}$$

Another important property is that at the lowest electron energy served by the SDCP USSR the sum of spin vector rotation angles in all BMs is always equal to $90^0$:

$$\sum_{k=1}^{N} \alpha_{k,\min} = \frac{\pi}{2} \tag{3.10}$$

Moreover, $\cos\theta_{f,k}$, $\cos(\alpha_k - \theta_{i,k})$, and $\cos(\alpha_k + \theta_{i,k})$ are anti-symmetric functions with respect to the line $\alpha_k = \pi$ for $0 \leq \alpha_k \leq 2\pi$, $k = 1...N-1$. As the result, the spin rotation angles in USSR solenoids become symmetric functions with respect to the same line $\alpha_k = \pi$.

All this makes the spin rotator data analysis absolutely transparent and immediately defines the continuous electron energy ratio that can be covered by the SDCP USSR:

$$(\Delta E_{USSR,N})_r = 4N - 3 \qquad (3.11)$$

## 4. USSR SOLENOIDS

The operations of solenoids belonging to the USSR cell number $k$ have to follow the model (2.4). As a consequence, the cumulative functionality of solenoids in the whole USSR is described by the next equation system $(k = 1,...N)$:

$$\sin\varphi_{1,k} = \frac{\cos\theta_{f,k} - \cos\alpha_k \cos\theta_{i,k}}{\sin\alpha_k \sin\theta_{i,k}}$$
$$\sin(\varphi_{2,k} + \chi_k) = 0 \qquad (4.1)$$
$$\tan\chi_k = \frac{-\sin\alpha_k \cos\theta_{i,k} + \cos\alpha_k \sin\varphi_{1,k} \sin\theta_{i,k}}{\cos\varphi_{1,k} \sin\theta_{i,k}}$$

and keeping in mind all special cases associated with the condition $\sin\alpha_k \sin\theta_{i,k} = 0$ mentioned above.

Remind that the spin rotation angle in the bending magnet of the last USSR cell $(k = N)$ must match the angle that the spin vector has with the axis Oy at the entrance into this cell:

$$\theta_{i,N} = \theta_{f,N-1} = -\alpha_N \qquad (4.2)$$

This leads us to the fact that the solenoid at the entrance into the last USSR cell must always rotate electron spin by $90^0$ clockwise

$$\varphi_{1,N} = -\frac{\pi}{2} \qquad (4.3)$$

and $\varphi_{2,N}$ can take any value. Therefore, the solenoid placed at the end of the last USSR cell can be, for example, turned off or used to control the spin tune in the JLEIC electron machine.

Another interesting property of USSR structures directly follows from their definition. If, at some particular electron energy $E_{M,USSR}$, the sum of spin rotation angles in all BMs starting from cell number $N_l$ $(l = 1,...N-1)$ up to the last cell can be expressed as $\sum_{i=N_l}^{N}\alpha_i = \frac{\pi}{2} + \pi \cdot m$, where $m$ is a positive integer number, then there exists a special USSR solution. This solution is characterized by the next parameters. The solenoid at the entrance into cell number $N_l$ rotates the spin vector through the angle $\varphi_{1,N_l} = -\frac{\pi}{2}$ or $\varphi_{1,N_l} = \frac{\pi}{2}$ depending on whether $m$ is even or odd respectively. All the rest solenoids in those cells, except one at the exit from the USSR, which is always exceptional, must be switched off.

The electron energies, for which such solutions exist, are called "magic USSR energies". Likewise, the solutions themselves are called "magic USSR solenoid setups". It is easy to realize that at magic USSR energies all USSR cells from up to the end act as one single "macro cell".

We note that by definition the real electron spin rotation angles in solenoids, which are required to set up between the bending magnets belonging to USSR cells number $k-1$ and $k$, are the sums of the values of $\varphi_{1,k}$ and $\varphi_{2,k-1}$ defined by the system (4.1) above. If we denote such angles as $\Phi_{k-1,k}$ then they have to fulfill the next requirements:

$$\Phi_1 = \varphi_{1,1}$$
$$\Phi_{k-1,k} = \varphi_{1,k} + \varphi_{2,k-1} \quad (2 \leq k \leq N) \quad (4.4)$$
$$\Phi_N = \varphi_{2,N}$$

Here $\Phi_1$ and $\Phi_N$ are the spin rotation angles in solenoids located before the BM in cell number 1 and after the BM in cell number $N$ respectively.

Consider now a couple of USSR structures, which are the most practical for the JLEIC conditions.

## 5. USSR WITH TWO CELLS

If an SDCP USSR consists of only two cells $(N = 2)$, then we consistently get

$$\alpha_2 = \frac{1}{2} \cdot \alpha_1, \quad \alpha_{1,\min} = \frac{\pi}{3}, \quad \alpha_{1,\max} = \frac{5\pi}{3} \quad (5.1)$$

$$(\Delta E_1)_r = \frac{\alpha_{1,\max}}{\alpha_{1,\min}} = (\Delta E_2)_r = \frac{\alpha_{2,\max}}{\alpha_{2,\min}} \equiv (\Delta E_{USSR,2})_r = 5$$

So, if we want, for example, this rotator to function starting from the minimum electron energy

$$E_{\min} = 3 \text{ GeV}$$

then it will continuously cover the entire energy range from 3 to 15 GeV.

With the use of formula (2.1.1) and well-known relation

$$a\gamma = \frac{E(GeV)}{0.44065} \qquad (5.2)$$

where $E(GeV)$ is the electron energy in GeV, one gets that the electron momentum $(\hat{\alpha})$ and spin $(\alpha)$ rotation angles in a bending magnet relate to each other as

$$\hat{\alpha} = \frac{0.44065}{E(GeV)} \cdot \alpha \qquad (5.3)$$

Now, keeping in mind that at the minimum (3 GeV) electron energy

$$\alpha_1 + \alpha_2 = 3 \cdot \alpha_2 = \frac{\pi}{2}$$

we obtain the required strengths of the USSR bending magnets

$$\hat{\alpha}_1 = \frac{0.44065 \cdot \pi}{9} \cong 0.049 \cdot \pi = 0.1539 \Rightarrow 8.8^0 \qquad (5.4)$$

$$\hat{\alpha}_2 = \frac{\hat{\alpha}_1}{2} \cong 4.4^0 \qquad (5.5)$$

The working parameters of the USSR for some electron energies are presented in Table 5.1 below. We note that this USSR is nothing more than the USR that was presented in all our previous JLEIC related publications (see paper [2], for example).

| Electron energy (GeV) | $\alpha_1$ | $\alpha_2$ | $\Phi_1$ | $\Phi_{12}$ |
|---|---|---|---|---|
| 3 | $60^0$ | $30^0$ | $-90^0$ | 0 |
| 5 | $100^0$ | $50^0$ | $-40.75^0$ | $-98.51^0$ |
| 6 | $120^0$ | $60^0$ | $-35.26^0$ | $-109.47^0$ |
| 9 | $180^0$ | $90^0$ | $90^0$ * | 0 * |
| | | | 0 * | $-90^0$ * |
| 11 | $220^0$ | $110^0$ | $-32.15^0$ | $-115.71^0$ |

Table 5.1: Electron spin vector rotation angles in an SDCP USSR consisting of two cells. Here $\alpha_1$ and $\alpha_2$ are spin rotations in bending magnets, $\Phi_1$ and $\Phi_{12}$ in solenoids at the entrance into the USSR and between cells respectively. The rotation angle $\Phi_2$ in the solenoid at the exit from the USSR can be any value. One sees that at 9 GeV there exist two "magic USSR solenoid setups", which are denoted by the asterisk (*) sign.

## 6. USSR WITH THREE CELLS

If an SDCP USSR consists of three cells $(N = 3)$ and operates starting from the same minimum energy $E_{min} = 3$ $GeV$ then it will continuously cover the electron energy range from 3 to 27 GeV.

From the next obvious relations

$$\alpha_1 = \alpha_2 = 2 \cdot \alpha_3 \ , \ \alpha_{1,min} = \frac{\pi}{5} \ , \ \alpha_{1,max} = \frac{9\pi}{5} \tag{6.1}$$

one easily obtains the required strengths of USSR bending magnets

$$\hat{\alpha}_1 = \hat{\alpha}_2 \cong 5.28^0 \ , \qquad \hat{\alpha}_3 \cong 2.64^0 \tag{6.2}$$

The USSR operational parameters for some particular electron energies are shown in Table 6.1 below.

| Electron energy (GeV) | $\alpha_1$ | $\alpha_2$ | $\alpha_3$ | $\Phi_1$ | $\Phi_{12}$ | $\Phi_{23}$ |
|---|---|---|---|---|---|---|
| 3 | $36^0$ | $36^0$ | $18^0$ | $-90^0$ | 0 | 0 |
| 5 | $60^0$ | $60^0$ | $30^0$ | $-35.26^0$ | $-35.75^0$ | $-98.9^0$ |
|   |        |        |        | 0 *       | $-90^0$ *  | 0 *        |
| 7 | $84^0$ | $84^0$ | $42^0$ | $-24.14^0$ | $-47.78^0$ | $-119.63^0$ |
| 9 | $108^0$ | $108^0$ | $54^0$ | $-18.96^0$ | $-55.12^0$ | $-129.62^0$ |
|   |         |         |        | $90^0$ *   | 0 *        | 0 *         |
| 12 | $144^0$ | $144^0$ | $72^0$ | $-15.43^0$ | $-61.2^0$ | $-136.64^0$ |
| 15 | $180^0$ | $180^0$ | $90^0$ | $-90^0$ * | 0 * | 0 * |
|    |         |         |       | 0 *       | $90^0$ * | 0 * |
|    |         |         |       | 0 *       | 0 *      | $-90^0$ * |

Table 6.1: Electron spin vector rotation angles in an SDCP USSR consisting of three cells. One notices that "magic USSR solenoid setups", which are marked by the asterisk (*) sign, exist at 5, 9, and 15 GeV.

## 7. USSR SPIN ROTATION MATRIX AND SPIN TUNE

Return now to the cumulative spin rotation matrix of a USSR cell defined by the equation (2.3):

$$M_c = \begin{pmatrix} \cos\alpha \cos\varphi_2 \cos\varphi_1 - \sin\varphi_2 \sin\varphi_1 & -\sin\alpha \cos\varphi_2 & \cos\alpha \cos\varphi_2 \sin\varphi_1 + \sin\varphi_2 \cos\varphi_1 \\ \sin\alpha \cos\varphi_1 & \cos\alpha & \sin\alpha \sin\varphi_1 \\ -\cos\alpha \sin\varphi_2 \cos\varphi_1 - \cos\varphi_2 \sin\varphi_1 & \sin\alpha \sin\varphi_2 & -\cos\alpha \sin\varphi_2 \sin\varphi_1 + \cos\varphi_2 \cos\varphi_1 \end{pmatrix}$$

Acting analogously to what was done in Chapter 2, denote the angle that the electron spin vector has with the axis Oy at the entrance into the cell as $\theta_i$ and the same angle after the cell as $\theta_f$. Formulate and prove the next Proposition.

**Proposition 7.1**

The USSR cell spin rotation matrix $M_c$ is a superposition of three consecutive rotations:

- by the angle $-\theta_i$ around the radial axis Ox,

- by some angle $\beta$, which is uniquely determined from the cell parameters, around the longitudinal axis Oy

and

- by the angle $\theta_f$ around the radial axis Ox again.

Proof.

If we denote

$$B(\beta) \equiv ROT(Oy, \beta) = \begin{pmatrix} \cos\beta & 0 & \sin\beta \\ 0 & 1 & 0 \\ -\sin\beta & 0 & \cos\beta \end{pmatrix} \qquad (7.1)$$

and

$$X(\theta) \equiv ROT(Ox, \theta) = \begin{pmatrix} 1 & 0 & 0 \\ 0 & \cos\theta & -\sin\theta \\ 0 & \sin\theta & \cos\theta \end{pmatrix} \qquad (7.2)$$

then we have to prove that

$$M_c = X(\theta_f) \cdot B(\beta) \cdot X(-\theta_i) \qquad (7.3)$$

or

$$X(\theta_f) \cdot B(\beta) = M_c \cdot X(\theta_i) \qquad (7.4)$$

Write the last expression in its explicit form:

$$\begin{pmatrix} \cos\beta & 0 & \sin\beta \\ \sin\beta\cdot\sin\theta_f & \cos\theta_f & -\cos\beta\cdot\sin\theta_f \\ -\sin\beta\cdot\cos\theta_f & \sin\theta_f & \cos\beta\cdot\cos\theta_f \end{pmatrix} = \quad (7.5)$$

$$\begin{pmatrix} \cos\alpha\cos\varphi_2\cos\varphi_1 - \sin\varphi_2\sin\varphi_1 & -\sin\alpha\cos\varphi_2 & \cos\alpha\cos\varphi_2\sin\varphi_1 + \sin\varphi_2\cos\varphi_1 \\ \sin\alpha\cos\varphi_1 & \cos\alpha & \sin\alpha\sin\varphi_1 \\ -\cos\alpha\sin\varphi_2\cos\varphi_1 - \cos\varphi_2\sin\varphi_1 & \sin\alpha\sin\varphi_2 & -\cos\alpha\sin\varphi_2\sin\varphi_1 + \cos\varphi_2\cos\varphi_1 \end{pmatrix} \cdot \begin{pmatrix} 1 & 0 & 0 \\ 0 & \cos\theta_i & -\sin\theta_i \\ 0 & \sin\theta_i & \cos\theta_i \end{pmatrix}$$

We see that the second columns of the matrices $X(\theta_f)\cdot B(\beta)$ and $X(\theta_i)$ are connected by the relation, which is nothing more than the basic USSR cell equation system (2.4). The solutions of this system are known from the previous chapters of this paper.

Now, with these solutions in hands, let's take a look at the first row of the matrix $X(\theta_f)\cdot B(\beta)$. Here we immediately notice that the angle $\beta$ is, indeed, uniquely determined.

The last statement concludes the proof.

What happens if one combines $N$ USSR cells into one spin manipulation structure? The answer is given by the second proposition.

**Proposition 7.2**

The cumulative USSR spin rotation matrix $M_{USSR,N}$ represents a superposition of two consecutive rotations:

- by the angle $\dfrac{\pi}{2}$ around the radial axis Ox

and

- by some angle $\tilde{\beta}_N$, which is uniquely determined from the USSR parameters,

   around the longitudinal axis Oy.

Proof.

Rewrite the above equation (7.3) for the USSR cell number $k$ as

$$M_{c,k} = X(\theta_{f,k}) \cdot B(\beta_k) \cdot X(-\theta_{i,k}) \tag{7.6}$$

Now keeping in mind that $\theta_{i,1} = -\dfrac{\pi}{2}$, $\theta_{i,k+1} = \theta_{f,k}$, and $\theta_{f,N} = 0$ calculate the cumulative USSR spin rotation matrix:

$$M_{USSR,N} = \prod_{k=1}^{N} X(\theta_{f,k}) \cdot B(\beta_k) \cdot X(-\theta_{i,k}) = \tag{7.7}$$

$$= X(\theta_{f,N}) \cdot \prod_{k=1}^{N} B(\beta_k) \cdot X(-\theta_{i,1}) = \prod_{k=1}^{N} B(\beta_k) \cdot X\left(\frac{\pi}{2}\right) = B(\tilde{\beta}_N) \cdot X\left(\frac{\pi}{2}\right)$$

The last expression concludes the proof.

It is not difficult to realize that two propositions of this chapter allow one to determine the spin tune of any JLEIC electron polarization manipulating setup.

## 8. PRACTICAL USSR TYPES AND RELATIONS BETWEEN THEM

In previous chapters, we have defined basic parameters of the SDCP USSR(d, +, p) electron spin rotator type. We call these parameters fundamental USSR settings. These settings combined with obvious USSR symmetry properties that were mentioned in this paper above allow one to easily define the settings of other electron spin rotator types, which are important for JLEIC conditions.

In particular, in case of USSR(d, -, p), all USSR fundamental settings are valid if we substitute $\alpha_k$ by $-\alpha_k$.

Another example is USSR(u, +, p), which works exclusively in the [yz, U] half plane. The USSR(d, +, p) results are valid for this rotator if one substitutes $\theta_{f,k}$ by $-\theta_{f,k}$.

We note that due to their (xy)-mirror symmetry, cells of different USSR(I1,I2,I3) types can always be combined in "hybrid" USSR structures, which allow the spin vector motion between [yz, D] and [yz, U] half planes. In some conditions, such "hybrid" structures can significantly increase the efficiency of the used spin manipulating system.

## 9. USSR IN CASE OF VERTICAL BENDS AT THE ENDS OF RING ARCS

In our original spin manipulation design paper [2], we assumed that the EIC electron ring might have vertical bending magnets at the ends of the machine arcs. The absolute value of the bending angle of such magnets $|\hat{\alpha}_V|$ was supposed to be about 0.022 radians or 1.26°.

When the electron has the energy smaller than 60 GeV then the absolute value of its spin rotation angle in each of these magnets $|\alpha_V|$ is less than π. In this case (otherwise, one should obviously divide this magnet in two or more parts), in accordance with what was described in this paper above, one can consider such a magnet as a part of the last SDCP USSR cell and the electron spin manipulation properties are determined by the next Proposition.

**Proposition 9.1**

If the JLEIC electron ring has vertical dipoles with the absolute value of the bending angle $|\hat{\alpha}_V|$ and if the range of the electron energies limits the electron spin rotation angles in such dipoles by the relation $|\alpha_V| < \pi$, then in the sense of electron spin manipulation functionality, the last cell of the SDCP USSR(d, +, p) is equivalent to a structure, which includes:

1. the same solenoid as at the entrance of the cell;
2. a horizontal (arc) dipole magnet with its bending angle

$$\hat{\alpha}_{N,H} = \hat{\alpha}_N - |\hat{\alpha}_V| \; ; \tag{9.1}$$

3. an additional solenoid rotating electron spin by the angle $\mp \dfrac{\pi}{2}$ depending on whether the value of $\hat{\alpha}_V$ is correspondingly negative or positive;
4. the existing vertical dipole magnet bending electron beams by the angle $\hat{\alpha}_V$ ;
5. another additional solenoid rotating electron spin by the angle $\pm \dfrac{\pi}{2}$ depending on whether the value of $\hat{\alpha}_V$ is correspondingly negative or positive;
6. the same solenoid as at the exit of the cell.

The proof of this Proposition immediately follows from the next two obvious expressions:

$$ROT(Oy, \frac{\pi}{2}) \cdot ROT(Ox, \alpha_v) \cdot ROT(Oy, -\frac{\pi}{2}) \cdot ROT(Oz, \alpha_h) = ROT(Oz, \alpha_h - \alpha_v)$$

and (9.2)

$$ROT(Oy, -\frac{\pi}{2}) \cdot ROT(Ox, \alpha_v) \cdot ROT(Oy, \frac{\pi}{2}) \cdot ROT(Oz, \alpha_h) = ROT(Oz, \alpha_h + \alpha_v)$$

The results of this chapter allow one to easily apply spin manipulation techniques developed for the flat JLEIC electron ring to the case with vertical bends, which can be considered for various JLEIC beam colliding options.

We note that other possibilities to create USSR structures suitable for the JLEIC electron ring with vertical bends directly follow from the USSR cell (xy)-mirror symmetry.

## CONCLUSIONS

The results presented in this paper allow one to design and build a variety of electron spin transformation structures, which can be described by simple linear mathematical models. Being integrated into storage ring optics, such structures become operational in large continuous beam energy intervals and extremely efficient for nuclear physics experiments.